\documentclass[11pt,a4paper]{article}
\usepackage{jcappub}
\usepackage[top=1in, bottom=1in, left=1in, right=1in]{geometry}
\usepackage{amsmath, amssymb, amsfonts, mathrsfs, graphicx, fancyhdr}
\usepackage[singlelinecheck=false]{caption}
\usepackage{subcaption}
\usepackage{setspace}
\usepackage{color}
\usepackage{epstopdf}
\definecolor{darkgreen}{rgb}{0,0.5,0}
\definecolor{darkblue}{rgb}{0,0,0.5}
\usepackage[usenames,dvipsnames]{xcolor}
\usepackage{bbm}
\usepackage{booktabs}
\newcommand{\be}{\begin{equation}}
\def\ee{\end{equation}}

\def\bea{\begin{eqnarray}}
\def\eea{\end{eqnarray}}

\newcommand{\ba}{\begin{eqnarray}}
\newcommand{\ea}{\end{eqnarray}}
\newcommand{\mpl}{m_{\rm Pl}}

\title{ $R^2\log R$ quantum corrections and the inflationary observables}

\begin{abstract}
{We study a model of inflation with terms quadratic and logarithmic in the Ricci scalar, where the gravitational action is $f(R)=R+\alpha R^2+\beta R^2 \ln R$. 
These terms are expected to arise from one loop corrections involving matter fields in curved space-time. 
  The spectral index $n_s$ and the tensor to scalar ratio yield $10^{-4}\lesssim r\lesssim0.03$ and  $0.94\lesssim n_s \lesssim 0.99$. i.e. $r$ is an order of magnitude bigger or smaller than the original Starobinsky model which predicted $r\sim 10^{-3}$. Further enhancement of $r$ gives a scale invariant $n_s\sim 1$ or higher. 
 Other inflationary observables are $d n_s/d\ln k \gtrsim -5.2 \times 10^{-4},\, \mu \lesssim 2.1 \times 10^{-8} ,\, y \lesssim 2.6 \times 10^{-9}$. Despite the enhancement in $r$, if the recent BICEP2 measurement stands, this model is disfavoured.} 
\end{abstract}

\author{Ido Ben-Dayan$^1$, Shenglin Jing$^{2,3}$,
Mahdi Torabian$^4$,
Alexander Westphal$^1$, 
Lucila Zarate$^5$}
\affiliation{$^1$DESY Theory Group, Notkestrasse 85, D-22607 Hamburg, Germany}
\affiliation{$^2$Canadian Institute for Theoretical Astrophysics, University of Toronto, 60 St.George Street, Toronto, ON, M5S 3H8, Canada}
\affiliation{$^3$Department of Astronomy and Astrophysics, University of Toronto, 50 St.George Street, Toronto, ON, M5S 3H4, Canada}
\affiliation{$^4$School of Particles and Accelerators, Institute for Research in Fundamental Sciences, P.O.Box 19395-5531, Tehran, Iran}
\affiliation{$^5$Institut f\"ur Theoretische Physik, Universit\"at Hamburg, Notkestrasse 85,D- 22761 Hamburg, Germany}

\begin{document}
\hfill{DESY-14-064}

\maketitle

%
\section{Introduction} 
The first data released from the Planck satellite had great implications for cosmic inflation and constrained many inflationary scenarios \cite{Ade:2013uln}. Higher-curvature term driven inflation, {\it a.k.a.} Starobinsky model \cite{Starobinsky:1980te}, nicely lies in the center of the maximum likelihood contours of PLANCK  \cite{Ade:2013uln}.  
The Starobinsky models adds a term quadratic in the Ricci scalar to the Einstein-Hilbert action, which provides for a regime of slow-roll inflation, asymptoting to de Sitter space at large $R$. Although fourth order in derivatives, the model is ghost-free and unitary \cite{Stelle:1977}, see also \cite{Kofman, Vilenkin} for further analysis. The full action is conformally equivalent to the Einstein gravity plus a dynamical propagating scalar field with a scalar potential. For large field values, the potential approaches a plateau exponentially fast providing asymptotically an effective cosmological constant. However, this model is ruled out if the recent BICEP2 \cite{BICEP2} result stands.

Initially, this model was based on the observation that the one-loop effective action of quantum fields coupled to gravity contains higher order curvature terms. In a particular limit, these are given to leading order by terms quadratic and logarithmic in the Ricci scalar. In its original spirit however (computing the $R^2$ coefficient just from the SM sector), this setup failed to yield  observationally viable inflation because it produced too much curvature perturbations.
Originally, the Starobinsky model arose from two corrections to the Einstein-Hilbert action:
a vacuum polarization and a particle production effective term. The first provides a contribution with an
$R$-dependence of the form ($\tilde \alpha R^2 + b R^2 \ln(R/\mu)$) with $\mu$ the renormalization scale and $\tilde \alpha,b$
coefficients completely determined by the number and spin of the fields present in the theory.
The second contribution is quadratic in R ($\propto R^2$) with arbitrary proportionality constant.
The sum of both provides a Starobinsky setup with an effective action described by a function $f(R)$ given by $R+\alpha R^2 +\beta R^2 \ln R$,
in which $\alpha=(1-b\ln \mu)\tilde \alpha$ and $\beta=b$. What is concurrently called {\it the Starobinsky model} simply corresponds to $\alpha=1/6M^2,\, \beta \simeq 0$.
It is important to note that in this
 model, we choose $M$ appropriately to fix the correct magnitude of the primordial power spectrum, while the $R^2\log R$ term is generally discarded as being negligible.
This approximation is valid if we compute $\beta$ using just the Standard Model sector degrees of freedom.
However, if we consider gravity as an effective theory with usually additional degrees of freedom, such as e.g. in string theory with its many vacuum solutions,
$\beta$ becomes effectively an adjustable parameter.

Hence, in the present letter
we will study the observational implications generated by the introduction of such term. More precisely, we will treat it as a correction
and compute the bounds on $\beta$ to still have a phenomenologically viable cosmological model considering the present constraints from inflation. 
A potentially appealing property of the model at the classical level is the fact that for certain range of parameters the model pushes the Big Bang singularity to the infinite past $t\rightarrow -\infty$ \cite{Gurovich:1979xg}. This would be interesting, since inflationary models with a scalar field are past incomplete and reach the Big Bang singularity in a finite time \cite{Borde}. However, this original motivation for the Starobinsky model failed because of instabilities against quantum mechanical decay~\cite{Mukhanov:1981xt}.

The Starobinsky model has another important property that at large scalar field
 values~$\chi \gg 1$  , the potential approaches a constant exponentially, thus having an approximate shift symmetry at large values (in absence of the linear term $R$ the model has an exact shift symmetry). Therefore, deformations such as $R^n$, considered for example in \cite{Berkin} break the shift symmetry. The $\ln R$ correction we use here has the advantage of spoiling the shift symmetry useful for realizing inflation 
 only weakly.  

While this
 manuscript was prepared for publication,~\cite{Sannino} appeared analyzing deviations from the Starobinsky model of the form $f(R)=R+\alpha R^{2(1-\gamma)}m_{Pl}^{4\gamma}$. Our analysis corresponds to the "first order" case with $\gamma \ll 1$, which limits the amount of tensor modes to $r < 0.03$ for $n_s<1$. Achieving higher $r$ to better match the BICEP2 results~\cite{BICEP2} requires adding higher-order terms in $f(R)$. Adding these terms, or re-summing an infinite series of $R^n$ corrections into a functional form $\Delta f(R)$ usually requires UV information (the choice $\alpha R^{2(1-\gamma)}$ in~\cite{Sannino} was given without a UV embedding).
 

%
\section{Formulation} 
The starting point is to postulate the Jordan frame action in four spacetime dimensions
\be S = \frac{\mpl^2}{2} \int {\rm d}^4 x (-g)^{1/2} f(R),\ee 
where modifications to the Einstein-Hilbert term are encoded in 
\be f(R) = R + \alpha R^2 + \beta R^2  \ln R.\ee
The two phenomenological parameters $\alpha$, $\beta$ in the action will be fixed momentarily and we work in units of $\mpl=1$. 
 The Einstein frame action can be reached by a Weyl transformation of the metric
\be g^E_{\mu\nu} = e^{\tilde\chi}g_{\mu\nu},\ee
where the dimensionless Weyl scalar field is defined as
\ba e^{\tilde\chi} &\equiv& f'(R) =1 + (2\alpha+\beta)R\Big[1+\frac{2\beta}{2\alpha+\beta}\ln R\Big].\label{Weyl-Ricci}\ea
where the potential is given by
\be V = \mpl^2\frac{(\alpha+\beta) R^2}{2 f'^2}\Big(1+\frac{\beta}{\alpha+\beta}\ln R\Big)=\mpl^2\frac{(\alpha+\beta) R^2\Big(1+\frac{\beta}{\alpha+\beta}\ln R\Big)}{2\left(1 + (2\alpha+\beta)R\Big[1+\frac{2\beta}{2\alpha+\beta}\ln R\Big]\right)^2}. \ee
Taking the limit of $\beta \rightarrow 0$ reproduces the Starobinsky model.
Recovering dimensions and multiplicative factors for the dynamical field $\chi\equiv (3/2)^{1/2}\mpl\ \tilde\chi$, the action canonical in the metric and scalar field reads
\be S_E =  \int {\rm d}^4 x (-g_E)^{1/2} \bigg(\frac{\mpl^2}{2} R_E - \frac{1}{2}(\partial _\mu \chi)^2 -  V(\chi)\bigg),\ee 
The inversion of \eqref{Weyl-Ricci} gives $R$ in terms of the Lambert function, also called ProductLog, $W_k$:
\be
\label{Rexact}
 R =  \frac{\big(e^{\tilde\chi}-1\big)}{2\beta {\rm W}_{\rm k}(X)}, \quad X\equiv \frac{e^{\tilde \chi}-1}{2\beta}e^{(2\alpha+\beta)/2\beta}
 \ee 
where ${W_ k}$ is the Lambert function of branch ${\rm k}=0$ for $\beta>0$, and , ${\rm k}=-1$ when $\beta<0$ .\\
The exact form of the potential for the scalar  in the Einstein frame is
\be
V(\tilde \chi)=(1-e^{-\tilde \chi})^2\frac{1+2W_k(X)}{16 \beta W_k(X)^2}=\frac{(1-e^{-\tilde \chi})^2}{8\alpha}\frac{\alpha(1+2W_k(X))}{2 \beta W_k(X)^2}\equiv V_s\frac{\alpha(1+2W_k(X))}{2 \beta W_k(X)^2}
\ee
where it is evident that the inclusion of the quantum terms is a correction to the Starobinsky potential $V_s$.
All inflationary predictions can be readily derived from the above potential. However, for the sake of clarity, we will now simplify it considerably.
The approximations below work extremely well for $|\beta|\ll \alpha$. For larger $\beta$ one should use the exact potential. While in these cases r is enhanced for $\beta<0$, $n_s \gtrsim 1$ so we don't consider this case.

We work out the approximation as follows. The equation \eqref{Weyl-Ricci} can be inverted and solved for $R$ using the iterative method
\be 
R = \frac{e^{\tilde\chi}-1}{2\alpha+\beta+2\beta \ln R},
\ee
The zeroth level solution is the Starobinsky solution
\be 
R^{(0)} = \big(e^{\tilde \chi}-1\big)/(2\alpha),\ee the leading order reads
\be R^{(1)}= \frac{e^{\tilde\chi}-1}{2\alpha+\beta+2\beta \ln R^{(0)}}
\ee
..the n-th level
\be \ R^{(n)}= \frac{\big(e^{\tilde \chi}-1\big)}{2\alpha+\beta+2\beta \ln R^{(n-1)}}
\ee 
To conclude, $R^{(n)}$ expands the Lambert function in the limit $W_k(X)\gg 1$, i.e when $|\beta|\ll \alpha$, recovering thus \ref{Rexact}. In our analysis, it is accurate enough to consider the just leading order in R , the potential thus becomes
\be
\label{Vapprox}
V\simeq \frac{V_s}{1+\frac{\beta}{2\alpha}+\frac{\beta}{\alpha}\ln R^{(0)}}=\frac{V_s}{1+\frac{\beta}{2\alpha}+\frac{\beta}{\alpha}\ln\left[(e^{\tilde \chi}-1)/2\alpha\right]}.
\ee

%
\begin{figure}[h!]
\centering
\includegraphics[scale=0.5]{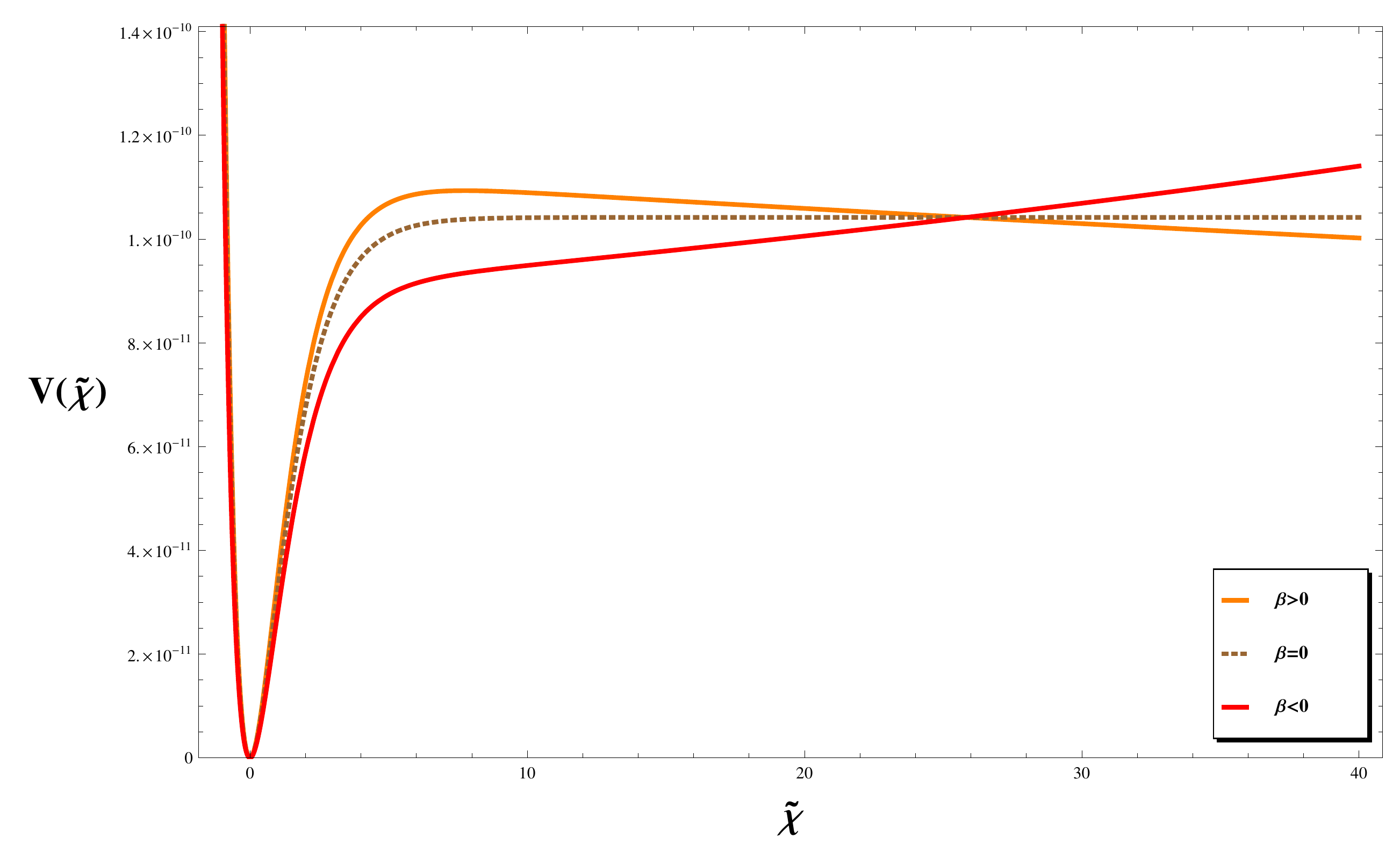}
\caption{{ The potential $V(\tilde \chi)$ for $\beta=0\,,\vert \beta \vert\sim 0.02 \alpha$ and $\alpha=1.2\times10^{-9}$. For $\beta>0$ the potential becomes unstable and has a runaway direction and a hill-top model, thus reducing $r$. For $\beta<0$ the potential tilts upwards, giving a larger slope thus enhancing $r$.}}
\label{fig:V_phi}
\end{figure}
The behaviour of the potential depends on the sign of the $\beta$ parameter as can be seen in Fig \ref{fig:V_phi}. For $\beta>0$ the potential becomes unstable with a runaway direction $\tilde\chi \rightarrow \infty$. However rolling towards the origin, one gets a hill-top type model. Typically hill-top models predict a red  $n_s$, in accordance with PLANCK measurements, with running $d{n_s}/d\ln k \gtrsim -5 \times 10^{-4}$; $r\lesssim 0.004$ and spectral distortions $\mu \lesssim 1.6 \times 10^{-8}$ , $y \lesssim 2.2 \times 10^{-9}$. 
In light of the recent BICEP2 detection of $r\in[0.12,0.2]$ we do not pursue further analysis of this case.
On the other hand, for $\beta<0$ the potential tilts upwards\footnote{It should be mentioned that the field range for which $f^{\prime}$ is positive, is limited in this case. Since the model is an effective description of a full UV theory we rely on the fact that this issue would be solved in the complete theory. Nevertheless, the inflationary phase is  always within the allowed region.}, potentially giving rise to enhanced tensor to scalar ratio $r$. 
We analyze the phenomenology of this scenario in the next section. 


%
\subsection{Inflationary Observables}
In order to discriminate models of inflation, the scalar power spectrum of curvature is parametrized as follows~\cite{Ade:2013uln}
\begin{equation}
\mathcal{P}(k) =\frac{1}{24 \pi^2} \frac{V}{\epsilon}=A_s \big(k/k_{\star} \big)^{n_s(k_{\star}) - 1 + \frac{1}{2} d n_s / d\ln{k} \ln{(k/k_{\star})} +...}\,,
\label{powerspectrumexpansion}
\end{equation}
where the subscript star indicates the pivot scale. For PLANCK it is $k_{\star} = 0.05$ Mpc$^{-1}$.  The amplitude $A_s$ of the scalar power spectrum is measured to be $A_s \simeq 2.2 \times 10^{-9}$. $n_s$ is the spectral index of the power spectrum. All CMB measurements show
 a nearly scale-invariant power spectrum, $n_s \lesssim 1$. The second-order contribution to the power spectrum is the running of spectral index ${\rm d}n_s/{\rm d}\ln k$, which measures how fast $n_s$ is changing across modes. 
In the case of slow-roll inflation, the following slow-roll parameters are defined to assist preliminary examinations of the model~\cite{Ade:2013uln}
\begin{equation}
\epsilon = \frac{1}{2}\frac{V'^2}{V^2}\,, \quad \eta = \frac{V''}{V}\,, \quad \xi^2 = \frac{V' V'''}{V^2}\,, 
\label{slowrollparameters}
\end{equation}
where a prime denotes derivation with respect to the scalar field $\phi$ for a potential $V(\phi)$. Notice that the second-order slow-roll parameter $\xi^2$ by convention can have any sign, and its superscript only indicates the order.

When the running of $n_s$ is insignificant, then $n_s$, $r$, and ${\rm d}n_s/{\rm d}\ln k$ can be expressed to first-order in terms of the above slow-roll parameters~\cite{Ade:2013uln},
\begin{align}
n_s & \approx 1 + 2 \eta - 6 \epsilon \label{ns} \,, \\
r & \approx 16 \epsilon \label{r} \,,\\
\frac{{\rm d}n_s}{{\rm d}\ln k} & \approx 16 \epsilon \eta - 24 \epsilon^2 - 2 \xi^2 \,. 
\end{align}

In this paper, all CMB observables are evaluated between $50$ and $60$ e-folds before the end of inflation, and we refer the point where we evaluate them to the CMB point and  inflation ends when $\epsilon = 1$.
The tensor-to-
scalar ratio, which is measuring the fractional power in primordial gravitational waves generated during inflation, directly relates to the scale of inflation. Moreover, in inflationary cosmology large values of $r$ are detectable via the B-mode polarization of the CMB.
 BICEP2~\cite{BICEP2} has recently claimed a detection of B-modes. If interpreted in the context of inflationary cosmology, their result corresponds to a tensor-to-scalar ratio with peak likelihood values $r\in[0.12,0.2]$ and $1-\sigma$ error of about $\pm 0.05$. 
According to PLANCK constraints \cite{Ade:2013uln}, the tensor to scalar ratio is restricted by $r<0.11$, the running of the spectral index by ${\rm d}n_s/{\rm d}\ln k=(-0.013\pm0.009)$ and the spectral index by $n_s=(0.961\pm0.007)$ at $95\%$CL. Hence, PLANCK is definitely in tension with the BICEP2 measurement. Alternatively, a recent analysis ~\cite{Spergel:2013rxa} suggested an increase in the spectral index to $n_s=(0.9671\pm0.0069)$, and a 2-sigma contour of $r$ which allows it to be as large as $r\sim 0.2$ thus relieving the tension between the BICEP2 and PLANCK measurements.

CMB and large scale structure (LSS) observations only probe a limited range of the scalar power spectrum, namely the wave-numbers ${H_0\lesssim k\lesssim 1\,\text{Mpc}^{-1}}$, which corresponds to the first $\sim\! \!8.4$ e-folds after $\phi_{CMB}$. 
However, constraints on smaller scales, therefore larger wave-numbers, play a crucial role in further constraining inflationary models. New types of observables have been proposed to serve this purpose \cite{Ben-Dayan:2013eza, Bringmann:2011ut, Li:2012qha}; in particular, the spectral distortions of the near-blackbody CMB spectrum~\cite{Chluba:2012we} can provide an important model independent constraint on the power spectrum at scales inaccessible by current CMB and LSS observations. 
 The spectral distortions depend on all quantities involved in parametrizing the primordial scalar power spectrum (\ref{powerspectrumexpansion}), at wave-numbers ${1\lesssim k \lesssim10^4 \,\text{Mpc}^{-1}}$, 
and are expected to be measured in near future experiments by PIXIE or PRISM,~ \cite{Kogut:2011xw, Andre:2013afa}. 
 For example, simple single-field models with a constant tilt $n_s=0.96$ without any higher order contributions would give $\mu\simeq 1.4 \cdot 10^{-8}$ and $y\simeq 3.3 \cdot 10^{-9}$. On the other hand, once included the running and running of running of the spectral index, enhancement of the power spectrum on the aforementioned scales will subsequently amplify the spectral distortions signal \footnote{A rather interesting possibility are models with $\Delta \phi \lesssim 1$ where detectable $r>0.01$ implies enhanced scale dependence of $n_s$ and in particular enhancement of power on small scales \cite{BenDayan:2009kv}. A further outcome of these models is enhanced spectral distortions. Such models could relieve the tension between PLANCK and BICEP2, so the measurement of spectral distortions could confirm or rule out these models.}. Following~\cite{Chluba:2012we}, one can compute approximately both types of spectral distortions as integrals over~$k$
\begin{align}
\mu &\approx 2.2 \, \int^\infty_{k_{min}} P_{\zeta}(k) \Bigg [ \textrm{exp} \bigg (-\frac{\hat k}{5400} \bigg ) - \textrm{exp} \bigg (- \bigg [ \frac{\hat k}{31.6} \bigg ] ^2 \bigg ) \Bigg ] \text{d}\ln{k} \nonumber\,, \\
y &\approx 0.4 \, \int^\infty_{k_{min}} P_{\zeta}(k) \, \textrm{exp} \Bigg (- \bigg [ \frac{\hat {k}}{31.6} \bigg ] ^2 \Bigg ) \text{d} \ln{k}\,,
\end{align}
where $P_{\zeta} (k) = 2 \pi^2 \mathcal{P}(k) / k^3$, $k_\text{min} \approx 1$ Mpc$^{-1}$, $\hat k = k$ Mpc.

%

\section{Phenomenology}

\begin{figure}
  \begin{minipage}[b]{.5\linewidth}
     \centering\includegraphics[scale=0.4]{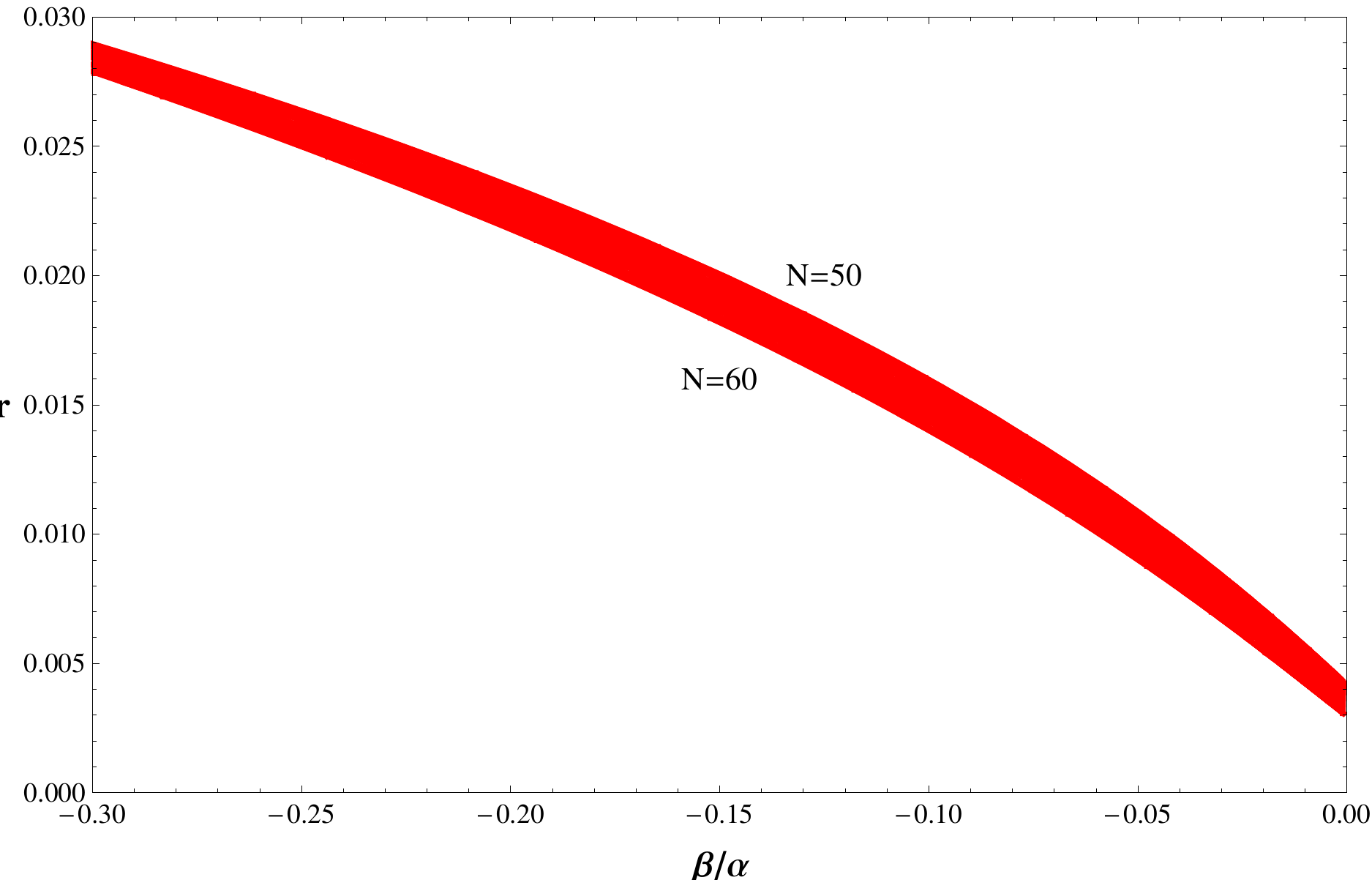}
  \end{minipage}%
  \begin{minipage}[b]{.5\linewidth}
     \centering\includegraphics[scale=0.405]{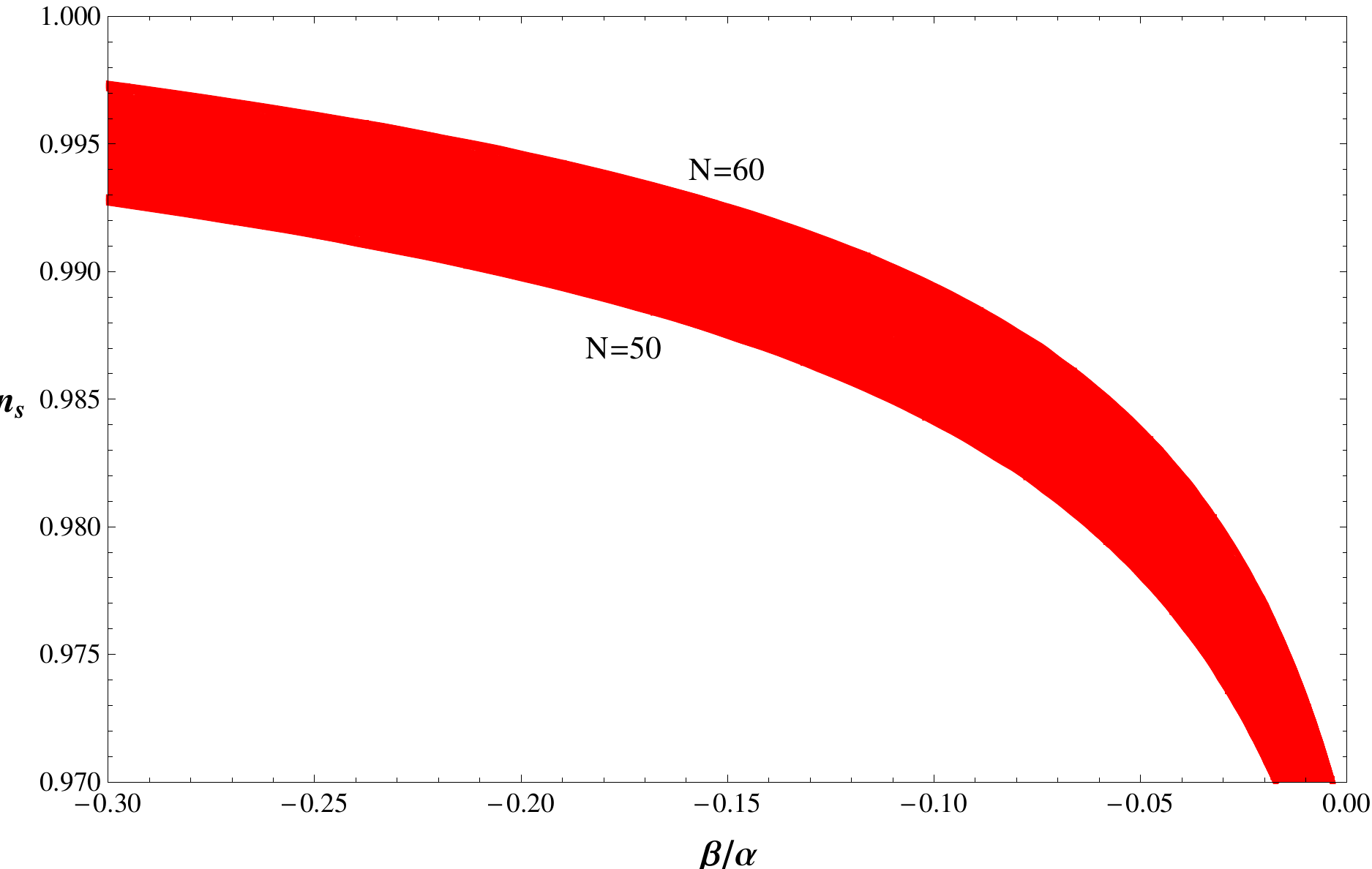}
  \end{minipage}
  \caption{$r$ vs. $\beta$ ( left ) and $n_s$ vs $\beta$ ( right ). The colored region corresponds to 50-60 e-folds of inflation.} 
\label{fig:observables}
\end{figure}

 Using the definitions in (\ref{slowrollparameters}) and the potential from (\ref{Vapprox}) we calculate the slow roll parameters as deviations from the Starobinsky model. These expressions are denoted by a subscript $s$. 
Defining $\delta=\frac{\beta/\alpha}{1+\frac{\beta}{2\alpha}+\frac{\beta}{\alpha}\ln\left[(e^{\tilde \chi}-1)/2\alpha\right]}$, the slow-roll parameters in terms of $\delta$ read (up to corrections of ${\cal O}(\partial_\chi^2(\ln R_s))={\cal O}(0.01)$ to the numerical coefficients): 
\begin{align}
\epsilon&=\epsilon_s-\delta\ \sqrt{\frac{4\epsilon_s}{3}}+\frac{\delta^2}{3} \\
\eta&=\eta_s-\delta\sqrt{\frac{16\epsilon_s}{3}}+\delta^2\frac{4}{3}\\
\xi^2&=\xi_s^2-\delta\ (2\sqrt{3\epsilon_s}\eta_s+\frac{\xi_s^2}{\sqrt{3\epsilon_s}})+\delta^2\ (8\epsilon_s+2\eta_s)+{\cal O}(\delta^3).
\end{align}
By using these relations in (\ref{ns}) and (\ref{r}) we obtained the observables
\begin{align}
 n_{s}&= (n_{s})_s+\delta \sqrt{\frac{16\epsilon_s}{3}}\\
 r&=r_s+16(-\delta\sqrt{\frac{4\epsilon_s}{3}}+\frac{\delta^2}{3})\\
\frac{{\rm d}n_s}{{\rm d}\ln k}  &=\left(\frac{{\rm d}n_s}{{\rm d}\ln k} \right)_s+\frac{1}{\sqrt 3} (32\epsilon_s^{3/2}-20\sqrt{\epsilon_s}\eta_s+2\xi_s^2/\sqrt{\epsilon_s})\cdot\delta+ \frac{4}{3}\eta_s\cdot \delta^2+{\cal O}(\delta^3)
\label{observables}
\end{align}
The number of e-folds in our model can be approximated by
\begin{equation}
N\simeq -\frac{3\log[\frac{1-\frac{\delta^2}{3\epsilon_s}}{1-\frac{\delta^2}{3}}]-6\text{ Arctanh}[\frac{\delta}{\sqrt{3\epsilon_s}}]}{\delta(2+\delta)}
\end{equation}
By demanding $N\sim 50-60$ and requiring the correct normalization of the power spectrum, we obtained the behaviour of the spectral index, the tensor to scalar ratio and the running as a function of $\beta$.
The left panel in Figure $2$ presents $r$ vs. $\beta/\alpha$. The right panel is $n_s$ vs. $\beta/\alpha$. The coloured region denotes the values corresponding to $50-60$ e-folds. For example $\beta/\alpha\sim -0.3$, $r\sim 0.025$, nearly an order of magnitude larger than the Starobinsky, $\beta=0$ prediction. However, this comes at a price of $n_s\simeq 0.995$ making it disfavoured even considering \cite{Spergel:2013rxa}.
In Figure $3$ we plot our predictions for $50-60$ e-folds in the $n_s-r$ plain on top of the BICEP2 contours (left) and the analysis of \cite{Spergel:2013rxa} (right).
Other predictions for $60$ e-folds are $d n_s/d\ln k \gtrsim -5.2 \times 10^{-4},\, \mu \lesssim 2.1 \times 10^{-8} ,\, y \lesssim 2.6 \times 10^{-9}$.

\begin{figure}
  \begin{minipage}[b]{.5\linewidth}
     \centering
\includegraphics[scale=0.45]{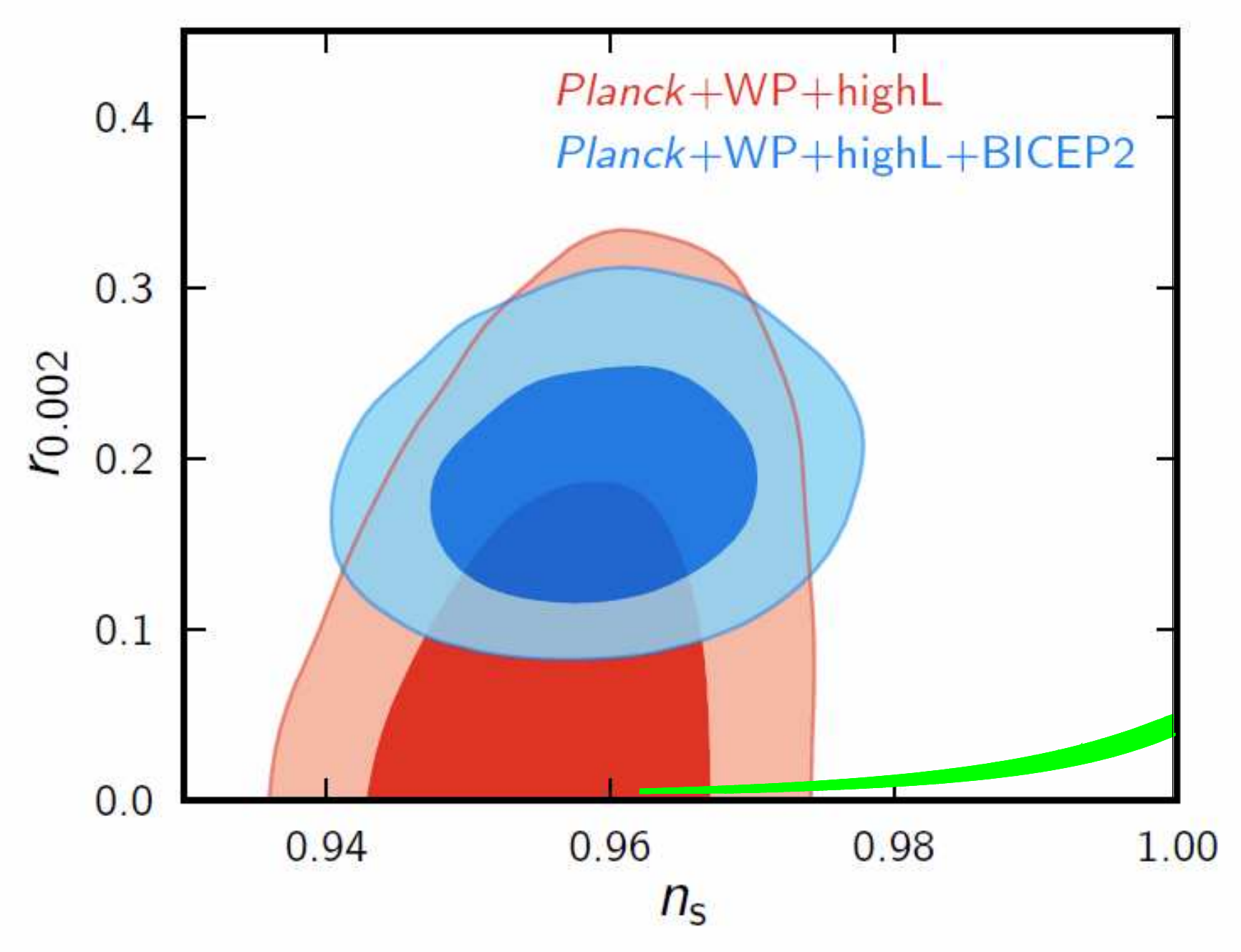}
   \vspace*{-5ex}
\label{fig:BICEP2}
  \end{minipage}%
   \begin{minipage}[b]{.5\linewidth}
 \centering
\includegraphics[scale=0.45]{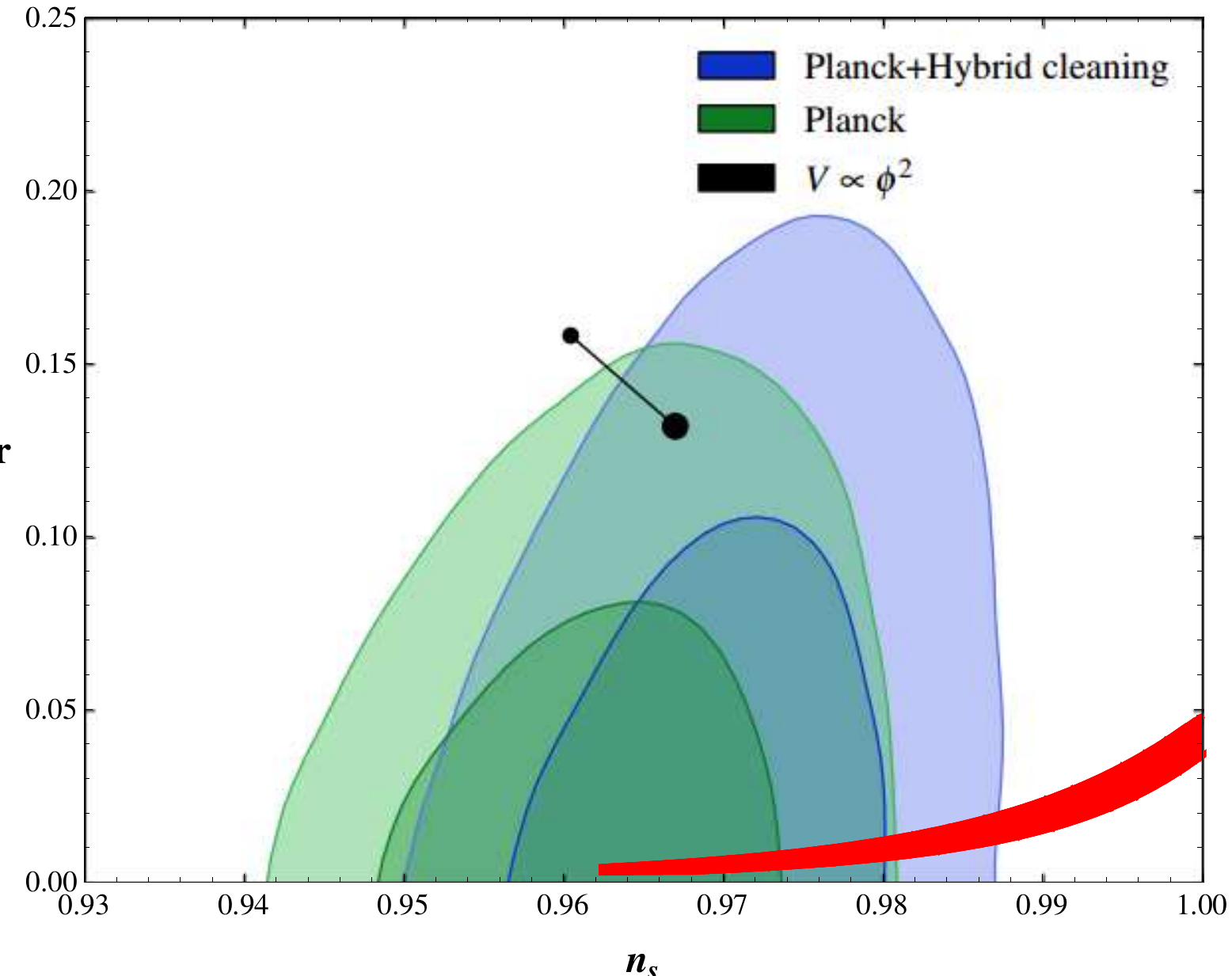}
  \end{minipage}
 \label{fig:spergel}
\caption{The predictions of the model corresponding to 50-60 efolds of inflation in the $r$ vs $n_s$ plane. In the left panel, we have overlaid our result (green region) on the BICEP2 $2 \sigma$ contours. As one can see, our result is beyond the $2 \sigma$ level of BICEP2 \cite{BICEP2}. On the right panel, we have overlaid our result (red region) on the PLANCK $2 \sigma$ contours as extended by \cite{Spergel:2013rxa}. The black line on the left indicates the predictions for the quadratic potential. As one can see, our result for $n_s$ barely hits the $2 \sigma$ level allowed by \cite{Spergel:2013rxa} analysis of PLANCK data when $r \gtrsim 0.02$.}
\end{figure}

\subsubsection*{Conclusions}
In this letter we analyzed the dynamics of inflation in a variant of the Starobinsky model specified by the function $f(R)=R+\alpha R^2+\beta R^2 \ln R$ in the light of PLANCK and BICEP2 data. This particular choice of $f(R)$ arises by taking into account quantum gravity corrections discussed e.g. in~\cite{Gurovich:1979xg}. While mostly taken to be small in earlier work, we have worked out the effects of the logarithmic correction here. Since we are ignorant about the full UV description of gravity, we regard the coefficient $\beta$ as arbitrary.   
Depending on the sign of $\beta$ we found that the predictions of the original Starobinsky model are significantly modified. 
The $\beta>0$ case induces a runaway direction as $\chi\rightarrow \infty$ and a hill-top model, now disfavoured by BICEP2.
For $\beta<0$, the tensor to scalar ratio, $r$, increases along with $n_s$. A tensor to scalar ratio of $r\sim 0.03$ yields $n_s\gtrsim 0.99$. Further enhancement of $r$ implies too large $n_s$ (unless higher-order terms beyond $R^2\ln\,R$ are added, as e.g. in~\cite{Sannino}).
Thus, when we combine the PLANCK and BICEP2 results they disfavor the $\beta<0$ modification as well.
%
\subsubsection*{Acknowledgments}
The work of  A.W. is supported by the Impuls und Vernetzungsfond of the Helmholtz Association of German Research Centres under grant HZ-NG-603. The work of I.B.-D. and L.Z. is supported by the German Science Foundation (DFG) within the Collaborative Research Center (CRC) 676 
``Particles, Strings,
 and the Early Universe''.
%
\bibliography{QuadLogR-vfinal}
\bibliographystyle{JHEP}

\end{document}